\RequirePackage{lineno}
\documentclass[aps,prl,twocolumn,showpacs,superscriptaddress,groupedaddress]{revtex4}  
\usepackage{graphicx}  
\usepackage{dcolumn}   
\usepackage{bm}        
\usepackage{amssymb}   
\usepackage{multirow}
\usepackage{epsfig}
\usepackage{wasysym}
\usepackage{booktabs}

\newcommand{\met}{\mbox{$\rlap{\kern0.15em/}E_T$}}

\begin{document}
\widetext
\hspace{5.2in} \mbox{FERMILAB-PUB-13-567-E}
\title{Erratum: Measurement of the $\boldsymbol{W}$ boson production charge asymmetry
in $\boldsymbol{p\bar{p}\rightarrow W+X \rightarrow e\nu +X}$ events at $\boldsymbol{\sqrt{s}=1.96}$~TeV}
%
\affiliation{LAFEX, Centro Brasileiro de Pesquisas F\'{i}sicas, Rio de Janeiro, Brazil}
\affiliation{Universidade do Estado do Rio de Janeiro, Rio de Janeiro, Brazil}
\affiliation{Universidade Federal do ABC, Santo Andr\'e, Brazil}
\affiliation{University of Science and Technology of China, Hefei, People's Republic of China}
\affiliation{Universidad de los Andes, Bogot\'a, Colombia}
\affiliation{Charles University, Faculty of Mathematics and Physics, Center for Particle Physics, Prague, Czech Republic}
\affiliation{Czech Technical University in Prague, Prague, Czech Republic}
\affiliation{Institute of Physics, Academy of Sciences of the Czech Republic, Prague, Czech Republic}
\affiliation{Universidad San Francisco de Quito, Quito, Ecuador}
\affiliation{LPC, Universit\'e Blaise Pascal, CNRS/IN2P3, Clermont, France}
\affiliation{LPSC, Universit\'e Joseph Fourier Grenoble 1, CNRS/IN2P3, Institut National Polytechnique de Grenoble, Grenoble, France}
\affiliation{CPPM, Aix-Marseille Universit\'e, CNRS/IN2P3, Marseille, France}
\affiliation{LAL, Universit\'e Paris-Sud, CNRS/IN2P3, Orsay, France}
\affiliation{LPNHE, Universit\'es Paris VI and VII, CNRS/IN2P3, Paris, France}
\affiliation{CEA, Irfu, SPP, Saclay, France}
\affiliation{IPHC, Universit\'e de Strasbourg, CNRS/IN2P3, Strasbourg, France}
\affiliation{IPNL, Universit\'e Lyon 1, CNRS/IN2P3, Villeurbanne, France and Universit\'e de Lyon, Lyon, France}
\affiliation{III. Physikalisches Institut A, RWTH Aachen University, Aachen, Germany}
\affiliation{Physikalisches Institut, Universit\"at Freiburg, Freiburg, Germany}
\affiliation{II. Physikalisches Institut, Georg-August-Universit\"at G\"ottingen, G\"ottingen, Germany}
\affiliation{Institut f\"ur Physik, Universit\"at Mainz, Mainz, Germany}
\affiliation{Ludwig-Maximilians-Universit\"at M\"unchen, M\"unchen, Germany}
\affiliation{Panjab University, Chandigarh, India}
\affiliation{Delhi University, Delhi, India}
\affiliation{Tata Institute of Fundamental Research, Mumbai, India}
\affiliation{University College Dublin, Dublin, Ireland}
\affiliation{Korea Detector Laboratory, Korea University, Seoul, Korea}
\affiliation{CINVESTAV, Mexico City, Mexico}
\affiliation{Nikhef, Science Park, Amsterdam, the Netherlands}
\affiliation{Radboud University Nijmegen, Nijmegen, the Netherlands}
\affiliation{Joint Institute for Nuclear Research, Dubna, Russia}
\affiliation{Institute for Theoretical and Experimental Physics, Moscow, Russia}
\affiliation{Moscow State University, Moscow, Russia}
\affiliation{Institute for High Energy Physics, Protvino, Russia}
\affiliation{Petersburg Nuclear Physics Institute, St. Petersburg, Russia}
\affiliation{Instituci\'{o} Catalana de Recerca i Estudis Avan\c{c}ats (ICREA) and Institut de F\'{i}sica d'Altes Energies (IFAE), Barcelona, Spain}
\affiliation{Uppsala University, Uppsala, Sweden}
\affiliation{Lancaster University, Lancaster LA1 4YB, United Kingdom}
\affiliation{Imperial College London, London SW7 2AZ, United Kingdom}
\affiliation{The University of Manchester, Manchester M13 9PL, United Kingdom}
\affiliation{University of Arizona, Tucson, Arizona 85721, USA}
\affiliation{University of California Riverside, Riverside, California 92521, USA}
\affiliation{Florida State University, Tallahassee, Florida 32306, USA}
\affiliation{Fermi National Accelerator Laboratory, Batavia, Illinois 60510, USA}
\affiliation{University of Illinois at Chicago, Chicago, Illinois 60607, USA}
\affiliation{Northern Illinois University, DeKalb, Illinois 60115, USA}
\affiliation{Northwestern University, Evanston, Illinois 60208, USA}
\affiliation{Indiana University, Bloomington, Indiana 47405, USA}
\affiliation{Purdue University Calumet, Hammond, Indiana 46323, USA}
\affiliation{University of Notre Dame, Notre Dame, Indiana 46556, USA}
\affiliation{Iowa State University, Ames, Iowa 50011, USA}
\affiliation{University of Kansas, Lawrence, Kansas 66045, USA}
\affiliation{Louisiana Tech University, Ruston, Louisiana 71272, USA}
\affiliation{Northeastern University, Boston, Massachusetts 02115, USA}
\affiliation{University of Michigan, Ann Arbor, Michigan 48109, USA}
\affiliation{Michigan State University, East Lansing, Michigan 48824, USA}
\affiliation{University of Mississippi, University, Mississippi 38677, USA}
\affiliation{University of Nebraska, Lincoln, Nebraska 68588, USA}
\affiliation{Rutgers University, Piscataway, New Jersey 08855, USA}
\affiliation{Princeton University, Princeton, New Jersey 08544, USA}
\affiliation{State University of New York, Buffalo, New York 14260, USA}
\affiliation{University of Rochester, Rochester, New York 14627, USA}
\affiliation{State University of New York, Stony Brook, New York 11794, USA}
\affiliation{Brookhaven National Laboratory, Upton, New York 11973, USA}
\affiliation{Langston University, Langston, Oklahoma 73050, USA}
\affiliation{University of Oklahoma, Norman, Oklahoma 73019, USA}
\affiliation{Oklahoma State University, Stillwater, Oklahoma 74078, USA}
\affiliation{Brown University, Providence, Rhode Island 02912, USA}
\affiliation{University of Texas, Arlington, Texas 76019, USA}
\affiliation{Southern Methodist University, Dallas, Texas 75275, USA}
\affiliation{Rice University, Houston, Texas 77005, USA}
\affiliation{University of Virginia, Charlottesville, Virginia 22904, USA}
\affiliation{University of Washington, Seattle, Washington 98195, USA}
\author{V.M.~Abazov} \affiliation{Joint Institute for Nuclear Research, Dubna, Russia}
\author{B.~Abbott} \affiliation{University of Oklahoma, Norman, Oklahoma 73019, USA}
\author{B.S.~Acharya} \affiliation{Tata Institute of Fundamental Research, Mumbai, India}
\author{M.~Adams} \affiliation{University of Illinois at Chicago, Chicago, Illinois 60607, USA}
\author{T.~Adams} \affiliation{Florida State University, Tallahassee, Florida 32306, USA}
\author{J.P.~Agnew} \affiliation{The University of Manchester, Manchester M13 9PL, United Kingdom}
\author{G.D.~Alexeev} \affiliation{Joint Institute for Nuclear Research, Dubna, Russia}
\author{G.~Alkhazov} \affiliation{Petersburg Nuclear Physics Institute, St. Petersburg, Russia}
\author{A.~Alton$^{a}$} \affiliation{University of Michigan, Ann Arbor, Michigan 48109, USA}
\author{A.~Askew} \affiliation{Florida State University, Tallahassee, Florida 32306, USA}
% ...
%
%
\collaboration{The D0 Collaboration\footnote{with visitors from
$^{a}$Augustana College, Sioux Falls, SD, USA,
$^{b}$The University of Liverpool, Liverpool, UK,
$^{c}$DESY, Hamburg, Germany,
$^{d}$Universidad Michoacana de San Nicolas de Hidalgo, Morelia, Mexico
$^{e}$SLAC, Menlo Park, CA, USA,
$^{f}$University College London, London, UK,
$^{g}$Centro de Investigacion en Computacion - IPN, Mexico City, Mexico,
$^{h}$Universidade Estadual Paulista, S\~ao Paulo, Brazil,
$^{i}$Karlsruher Institut f\"ur Technologie (KIT) - Steinbuch Centre for Computing (SCC),
D-76128 Karlsrue, Germany,
$^{j}$Office of Science, U.S. Department of Energy, Washington, D.C. 20585, USA
and
$^{k}$American Association for the Advancement of Science, Washington, D.C. 20005, USA.
}} \noaffiliation
\vskip 0.25cm

\date{December 10, 2013}
\pacs{13.38.Be, 13.85.Qk, 14.60.Cd, 14.70.Fm}

\maketitle
The measurement of the $W$ boson production charge asymmetry published in our recent Letter~\cite{d0_w}
employed a correction $K^{\pm}_{\text{eff}}$ to take into account the relative efficiency difference between electrons and positrons.
Based on a recent study~\cite{w_prd}, we realized that the determination of $K^{\pm}_{\text{eff}}$ was incorrect.
Instead of taking the ratio of the positron to electron efficiencies, we took the ratio of the numbers
of reconstructed positrons to electrons. In addition, we had not taken into account the solenoid
polarity when determining $K^{\pm}_{\text{eff}}$. These two problems have now been corrected.

The corrected $W$ boson charge asymmetry values measured using the updated
efficiency correction~\cite{w_prd} are given in Table~\ref{tab:wasymtab}.
These revised measurements, together with those from the CDF Collaboration~\cite{cdf_w}
are shown in Fig.~\ref{fig:compare_asym}.
The asymmetry values have changed relative to those in the original publication by $<2$\%,
with smaller asymmetry values for $|y_W|<0.6$ and larger asymmetry values for $0.8<|y_W|<2.4$,
compared to the published result~\cite{d0_w}.

\begin{figure}
\epsfig{file=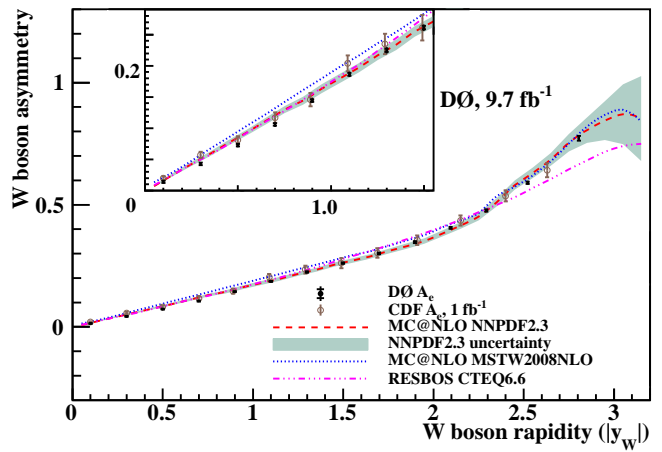, scale=0.48}
\caption{\small (color online). Measured $W$ boson charge asymmetry, after CP-folding,
compared to predictions and the CDF 1 fb$^{-1}$ result.
The points show the measured asymmetry, with the horizontal bars
delineating the
statistical uncertainty component and the vertical lines showing the total uncertainty.
The central value and uncertainty from {\sc mc@nlo}~\cite{mcatnlo} using the {\sc NNPDF2.3}~\cite{nnpdf} PDF sets and the predictions from
{\sc resbos}~\cite{resbos} using the CTEQ6.6~\cite{cteq} central PDF set and {\sc mc@nlo} using the MSTW2008NLO~\cite{mstw}
central PDF set are also shown.
The inset focuses on the $y_W$ region from 0 to 1.5.}
\label{fig:compare_asym}
\end{figure}

\begin{table}
\begin{center}
\caption{ CP-folded $W$ boson charge asymmetry for data and predictions
from {\sc mc@nlo} using the {\sc NNPDF2.3} PDFs tabulated in percent (\%)
for each $|y_W|$ bin. The $\langle|y_W|\rangle$ is calculated as the cross section weighted average of $y_W$ in each
bin from {\sc resbos} with {\sc photos}~\cite{photos}. For data, the first uncertainty is statistical and
the second is systematic.
The uncertainties on the prediction come from both the PDF uncertainties
and $\alpha_s$ uncertainties. }
\begin{tabular}{cccrr}
\hline
\hline
 Bin index &  $|y_W|$ &  $\langle|y_W|\rangle$ &   \multicolumn{1}{c}{Data} & Prediction      \\ \hline
\ \ 1&  0.0--0.2 &$0.10$ & $1.39 \pm 0.17 \pm 0.12$  &  $1.61 \pm 0.19$  \\
\ \ 2&  0.2--0.4 &$0.30$ & $4.28 \pm 0.18 \pm 0.19$  &  $5.06 \pm 0.33$ \\
\ \ 3&  0.4--0.6 &$0.50$ & $7.28 \pm 0.19 \pm 0.27$  &  $8.50 \pm 0.41$  \\
\ \ 4&  0.6--0.8 &$0.70$ & $10.59 \pm 0.20 \pm 0.30$  &  $12.05 \pm 0.53$  \\
\ \ 5&  0.8--1.0 &$0.90$ & $14.45 \pm 0.21 \pm 0.32$ &  $15.36 \pm 0.66$ \\
\ \ 6&  1.0--1.2 &$1.10$ & $18.63 \pm 0.22 \pm 0.39$ &  $18.86 \pm 0.74$ \\
\ \ 7&  1.2--1.4 &$1.30$ & $22.50 \pm 0.24 \pm 0.44$ &  $22.52 \pm 0.80$   \\
\ \ 8&  1.4--1.6 &$1.50$ & $26.12 \pm 0.27 \pm 0.42$ &  $26.30 \pm 0.85$   \\
\ \ 9& 1.6--1.8 &$1.70$ & $30.06 \pm 0.31 \pm 0.44$ &  $29.89 \pm 0.92$   \\
10 &  1.8--2.0 &$1.90$ & $34.73 \pm 0.35 \pm 0.49$   &  $34.04 \pm 1.08$  \\
11 &  2.0--2.2 &$2.10$ & $40.59 \pm 0.40 \pm 0.54$   &  $39.77 \pm 1.31$ \\
12 &  2.2--2.4 &$2.29$ & $47.65 \pm 0.44 \pm 0.56$   &  $47.73 \pm 1.62$  \\
13 &  2.4--2.7 &$2.52$ & $59.04 \pm 0.46 \pm 0.60$   &  $61.81 \pm 1.74$  \\
14 &  2.7--3.2 &$2.81$ & $77.24 \pm 0.93 \pm 0.66$   &  $78.05 \pm 4.36$  \\
\hline
\hline
\end{tabular}
\label{tab:wasymtab}
\end{center}
\end{table}

\clearpage

\makeatletter
\def\maketitle{%
\par\textbf{\large\bfseries\centering \@title}%
\par}

\makeatother

\title{Measurement of the $\boldsymbol{W}$ Boson Production Charge Asymmetry
in $\boldsymbol{p\bar{p}\rightarrow W+X \rightarrow e\nu +X}$ Events at $\boldsymbol{\sqrt{s}=1.96}$~TeV}
		
\maketitle

\begin{center}
We present a measurement of the $W$ boson production charge asymmetry 
in $p\bar{p}\rightarrow W+X \rightarrow e\nu +X$ events at a center of mass energy of 1.96 TeV,
using 9.7~fb$^{-1}$ of integrated luminosity collected with the D0 detector at the Fermilab Tevatron Collider.
The neutrino longitudinal momentum is determined by using a neutrino weighting method, 
and the asymmetry is measured as a function of the $W$ boson rapidity. 
The measurement extends over wider electron pseudorapidity region than
previous results and is the most precise to date, allowing for precise determination of
proton parton distribution functions in global fits.\\
(Dated: December 10, 2013)
\end{center}

\vspace{.2 in} 
At the Fermilab Tevatron Collider, production of $W^{\pm}$ bosons is dominated by the
annihilation of valence quarks in the proton ($u$, $d$) and antiproton
($\bar{d}$, $\bar{u}$).
The primary modes of production are $u+\bar{d}\rightarrow W^+$ and
$\bar{u}+d\rightarrow W^-$.
In the proton and antiproton, the $u$ ($\bar{u}$) quark generally carries more momentum than the $\bar{d}$ ($d$)
quark; thus, $W^+$ bosons are 
boosted in the proton direction and $W^-$ bosons in the antiproton direction~\cite{old_pdf1, old_pdf2, old_pdf3}.
The difference between $u$ and $d$ quark parton distribution functions (PDFs)
results in a charge asymmetry in the $W$ boson rapidity ($y_W$), defined as
\begin{eqnarray}
 A(y_W) = \frac{d\sigma_{W^+}/dy_W - d\sigma_{W^-}/dy_W} {d\sigma_{W^+}/dy_W + d\sigma_{W^-}/dy_W}.
\end{eqnarray}

Here, $d\sigma_{W^{\pm}}/dy_W$ is the differential cross section for $W^{\pm}$ boson
production, and $y_W$ is the $W^{\pm}$ boson rapidity, defined as
\begin{eqnarray}
y_W=\frac{1}{2}\ln\frac{E+p_z}{E-p_z},
\end{eqnarray}
where $E$ and $p_z$ are the energy and the longitudinal momentum, respectively, of the $W$ boson,
with the $z$ axis along the proton beam direction.

Previously published results include both lepton (from the $W$ boson
decay) and $W$ boson charge asymmetries.
The lepton charge asymmetry arises from the convolution of the $W$ boson asymmetry
and the $V-A$ structure of the $W$ boson decay. 
This implies that leptons at a specific 
rapidity originate from a wide range of $W$ rapidities,
and therefore from a wide range of parton $x$ values (where $x$ is the fraction of 
momentum of the proton carried by the parton), 
diluting the impact of these asymmetries when determining PDFs.
The lepton charge asymmetry in $W$ boson decays has been
measured by the CDF~\cite{CDF_results1, CDF_results2, CDF_results3}
and D0~\cite{d0_results_muon1, d0_results_em_old} Collaborations.
The latest lepton charge asymmetry measurement from the D0
Collaboration was performed in the $W\rightarrow \mu \nu$ muon channel by using data corresponding to 
7.3~fb$^{-1}$ of integrated luminosity~\cite{d0_muon}.
The lepton charge asymmetry has also been measured at the Large Hadron Collider (LHC) in $pp$ 
collisions by the ATLAS \cite{ATLAS_w} and CMS \cite{CMS_w}
Collaborations by using integrated luminosities of 0.03 and 0.84~fb$^{-1}$, respectively.
A direct measurement of the $W$ boson charge asymmetry was performed by using 
1~fb$^{-1}$ of integrated luminosity by the CDF~\cite{cdf_w} Collaboration.

The analysis presented in this Letter uses the $W\rightarrow e\nu$ decay mode
and employs the neutrino weighting method~\cite{cdf_method}.
In addition, this $W$ boson charge asymmetry analysis uses 10 times more integrated luminosity and
covers much larger rapidity range than the previous CDF result~\cite{cdf_w}. 
We use data corresponding to 9.7~fb$^{-1}$ of integrated luminosity~\cite{d0lumi} collected with the D0 detector~\cite{d0det0, d0det} 
between April 2002 and September 2011.
By extending the pseudorapidity coverage,
we can provide information about the PDFs for a broader range of $x$ ($0.002 < x < 0.99$ for 
electron pseudorapidity $|\eta^e| < $ 3.2~\cite{d0_coordinate}) at $Q^2\approx M^2_W$, where $Q^2$ is the 
squared momentum scale for the parton interactions and $M_W$ is the $W$ boson mass.
The $W$ boson charge asymmetry result places stringent constraints on 
the PDFs of valence quarks, which in turn will significantly reduce the
uncertainty on the measurements of $M_W$ and on other measurements at
the Tevatron and LHC.

\indent The D0 detector~\cite{d0det0, d0det} comprises a central tracking
system, a calorimeter, and a muon system. The central tracking 
system consists of a silicon microstrip tracker and a scintillating fiber tracker (CFT). 
The CFT provides coverage for charged particles 
at detector pseudorapidities of $|\eta_{\text{det}}|<1.7$.
Three liquid argon and uranium calorimeters provide coverage of
$|\eta_{\text{det}}|<3.5$ for electrons: the central calorimeter (CC) 
up to $|\eta_{\text{det}}|<1.1$ and two end calorimeters (EC) in the range
$1.5<|\eta_{\text{det}}|<3.5$. Gaps between the cryostats
create an inefficient electron detection region between $1.1<|\eta_{\text{det}}|<1.5$
that is excluded from the analysis.
Each calorimeter consists of an inner electromagnetic (EM) section, followed by hadronic sections. 

Events used in this analysis were collected with a set of calorimeter-based single-electron triggers.
To select $W\rightarrow e\nu$ events, we require one EM shower with
transverse energy will respect to the beam $25<E_T<100$~GeV measured in the calorimeter,
accompanied by large missing transverse energy of $\met>25$~GeV. 
$\met$ is estimated by the vector sum of the transverse components of
the energy deposited in the calorimeter ($u_T$) and the electron $E_T$.
An isolation requirement is imposed on the electron candidate, which is also required to
have a significant fraction of its energy deposited in the EM calorimeter,
compared to that deposited in the hadron calorimeter. 
Candidates in the CC must be in the range $|\eta_{\text{det}}|<1.1$, and
those in the EC must be within $1.5<|\eta_{\text{det}}|<3.2$, to allow
a precise measurement of electron energy.
The shower shape~\cite{shower_shape} must be
consistent with that expected for an electron, and the candidate is required to 
be spatially matched to a reconstructed track.
Because the CFT detector does not cover the entire $\eta_{\text{det}}$ region 
used in the analysis, electron selection criteria are separately defined
in four categories: CC electrons with full CFT coverage, EC electrons with full CFT coverage, 
EC electrons with partial CFT coverage, and EC electrons without CFT coverage.
Events are further required to have the reconstructed $p\bar{p}$ interaction vertex located within
40~cm of the detector center along the $z$ axis,
a reconstructed $W$ boson transverse mass ($M_T$) between 50 and 130~GeV,
where $M_T = \sqrt{2E_T\met (1-\cos\Delta \phi)}$, and $\Delta \phi$ is
the azimuthal angle between the electron and $\met$,
$u_T$ less than 60~GeV, and {\it SET} less than 250 or 500~GeV
depending on the data collection period,
where {\it SET} is the scalar sum of all transverse energies measured by the calorimeter
except those energies associated with electrons or with potential noise, 
reflecting the total activity in the event.

After applying the selection criteria described above, we retain 6~083~198 $W$ boson candidates. 
Of these, 4~466~735 are events with the electron in the CC region and 1~616~463 
with the electron in the EC region.
We have checked that the asymmetry results for $y_W>0$ are consistent with those for $y_W<0$, 
so we assume {\it CP} invariance---{\it i.e.}, $A(y_W)$ is equivalent to $-A(-y_W)$---and 
fold the data appropriately to increase the statistics in each $y_W$ bin.
The forward-backward charge asymmetries are measured in 14 bins of $y_W$
in the range $|y_W| < 3.2$. 
The bin widths are chosen by considering the sample size and the detector geometry
 to ensure that high $|y_W|$ bins retain sufficient statistics. 
 
Mismeasurement of the charge sign of the electron may result in 
a dilution of the $W$ boson charge asymmetry.
We measure the charge misidentification rate with $Z\rightarrow ee$ events,
using a ``tag-and-probe" method~\cite{tag-probe}. The tag electron must satisfy tight selection criteria to
ensure its charge is determined correctly. The charge misidentification rate varies
from $(0.18\pm0.01)$\% at $|\eta^e|=0$ to $(9.6\pm0.9)$\% at $|\eta^e|=3.0$,
where tracking momentum resolution is poor. 
The direction of the D0 solenoid magnetic fields was reversed
during data taking every two weeks on average, significantly reducing the 
charged particle reconstruction asymmetry in the detector; thus,
the charge misidentification rates of electrons and positrons are consistent for
different magnet polarities.
At $|\eta^e|=3.0$, the charge misidentification rates are $(9.4\pm1.3)$\%
for electrons and $(9.8\pm1.3)$\% for positrons
and are also consistent with each other at other $|\eta^e|$ values.

Monte Carlo (MC) samples for the $W\rightarrow e\nu$ process are generated 
by using the {\sc pythia}~\cite{pythia} event generator with {\sc CTEQ6L1} PDFs~\cite{cteq},
followed by a {\sc geant}-based simulation~\cite{geant} of the D0 detector. 
This simulation is then corrected for higher-order effects not included in {\sc pythia}.
The MC events are reweighted at the generator level in two dimensions 
($W$ boson transverse momentum, $p_T^W$, and $y_W$) to match {\sc resbos}~\cite{resbos} predictions.
To improve the accuracy of the MC detector simulation, 
further corrections are applied to the MC simulations including electron energy scale and resolution,
recoil system scale and resolution, selection efficiencies, trigger efficiencies,
instantaneous luminosity and {\it SET}, charge misidentification, and 
relative efficiency for identification of 
positrons and electrons ($K^{\pm}_{\text{eff}}$). These corrections are derived by comparing the
$Z\rightarrow ee$ data and {\sc pythia} MC distributions.
Because of imperfections in the modeling of the tracking detector, 
differences between the efficiency for electrons and positrons 
vary from 0.0\% at $|\eta^e|=0$ to 1\% at $|\eta^e|=3.0$.

The dominant source of background originates from 
multijet events, with one jet misreconstructed as an electron and with significant $\met$ due to the 
mismeasurement of the jet energy.
Smaller background contributions arise from other standard model (SM) processes and are
estimated by using {\sc pythia} MC samples normalized to the highest order available cross sections~\cite{nlo_xsection}.
These include $W\rightarrow \tau\nu$ events where the tau decays to an electron and neutrinos,
$Z\rightarrow ee$ events where one of the electrons is not identified, 
and $Z\rightarrow \tau \tau$ events with one tau decaying to
an electron and the other not identified.
The multijet background is estimated by using collider data by fitting the $M_T$
distribution in the region 50-130~GeV (after other SM backgrounds
have been subtracted) to the sum of the shape predicted by the
$W\rightarrow e\nu$ signal MC sample and the shape obtained from a multijet-enriched data sample. 
The multijet-enriched sample is selected by reversing the shower shape requirement on the electron candidates. 
The background contributions are determined as a function of $y_W$, and average contributions are 
4.0\% multijet events, 2.6\% $Z\rightarrow ee$, 2.2\% $W\rightarrow \tau\nu$, and 0.2\% $Z\rightarrow \tau \tau$.


In the determination of the longitudinal momentum of the neutrino ($p_z^{\nu}$)~\cite{cdf_method},
$M_W$ is fixed to the world average value of 80.385~GeV~\cite{wmass}.
The mass-energy relation constraint using the energy and momentum of the
neutrino and electron,
\begin{eqnarray}
M_W^2 = (E_e+E_{\nu})^2-(\vec{P_e}+\vec{P_{\nu}})^2,
\end{eqnarray}
implies that there are two solutions in $p_z^{\nu}$. The twofold
ambiguity can be partly resolved on a statistical basis
from the known $V-A$ decay distribution by using the 
decay angle between the electron and the proton ($\theta^*$) and 
from the $W^+$ and $W^-$ production cross sections as a function of $y_W$.
As expected, many off-shell $W$ boson decays do not satisfy the 
$M_W^2$ constraint. In this case, we obtain complex values for the $p_z^{\nu}$,
assume that the neutrino transverse momentum ($p_T^{\nu}$) is misreconstructed, 
and therefore scale $\met$ to the value for which
the imaginary part equals zero. This new $\met$ value is then used to
determine $p_T^{\nu}$ and therefore $y_W$.
To obtain the $W$ boson rapidity distributions, we assign different probabilities to the two $p_z^{\nu}$ solutions. 
This probability is related to the quark and antiquark $W^{\pm}$ boson production by
\begin{eqnarray}
 \label{eq:sea}
 \lefteqn{P_{\pm}\left(\cos\theta^*, y_W, p_T^W\right) =
  \left(1\mp\cos\theta^*\right)^2 +} \hspace{3.15in}\nonumber  \\
 Q\left(y_W, p_T^W\right)\left(1\pm\cos\theta^*\right)^2,~~~~~
\end{eqnarray}
where $P_{\pm}\left(\cos\theta^*, y_W, p_T^W\right) $ is the probability for $W$ boson 
production with a particular $\cos\theta^*$, $y_W$, and $p_T^W$. 
The first term in Eq.~(\ref{eq:sea}) represents the contribution from annihilation
with two quarks,
and the second term the contribution from annihilation with at least one
antiquark. 
The ratio $Q\left(y_W, p_T^W\right)$
between quark and antiquark $W$ boson production is a function of
$W$ boson rapidity and transverse momentum.
At the Tevatron, the $W$ boson production contribution from the antiquark and gluons is $\sim$
10\%.

Understanding the antiquark contribution is important for the 
asymmetry measurement, because $W$ bosons produced 
by antiquarks have opposite polarization from those produced by quarks.
The ratio of antiquark to quark $W$ boson production is determined 
by the angular distribution of $W\rightarrow e\nu$ decays.
We use the prediction of the fractions of antiquark to quark contributions 
from {\sc MC@NLO}~\cite{mcatnlo}, using the {\sc CTEQ6.6} PDF set, 
and parametrize the angular distributions as functions of $y_W$ and $p_T^W$,
using an empirical function to fit the ratio.

We use both $P_{\pm}$ and the differential cross section
$d \sigma^{\pm}_W / dy_W$ to define weights as in Eq.~(\ref{eq:winputs}).
The $W$ boson production cross section decreases in the forward region
due to the scarcity of high-$x$ quarks, and so solutions
leading to a central $W$ production are weighted more heavily than forward $W$
solutions. The weight factors $w_{i}$ for $W^+$ and $W^-$ are
\begin{eqnarray}
\label{eq:winputs}
   w^{\pm}_{i} = \frac{P_{\pm}\left(\cos\theta^*_{i}, y_{i},
p_T^W\right)d\sigma^{\pm}\left(y_{i}\right)/dy_W}{\sum\limits_{i}
P_{\pm}\left(\cos\theta^*_{i}, y_{i},
p_T^W\right)d\sigma^{\pm}\left(y_{i}\right)/dy_W},
\end{eqnarray}
where $i=1,2$ are the two solutions.
We use the predicted differential cross section $d\sigma^{\pm}_W/dy_W$ at 
next-to-next-to-leading order~\cite{vrap} as input when calculating the
weight factors for each neutrino $p_z^{\nu}$ solution.
We iterate by updating values of $d\sigma^{\pm}_W/dy_W$ to those obtained by using
the weight factor. This procedure converges after three or four iterations.

To measure the $W$ boson charge asymmetry, we apply unfolding corrections
to the measured $W^+$ and $W^-$ distributions to correct detection effects. 
The matrix inversion method~\cite{matrix} is used to correct for event migration effects. 
First, the product of acceptance and efficiency is applied to each bin to correct for the event selection
effects, and the $K_{\text{eff}}^{\pm}$ correction is used to equalize the efficiency response between
electrons and positrons. The migration matrices are obtained by using
the number of events in both the generator level $y_W$ bin $j$
and the reconstruction level $y_W$ bin $i$, 
divided by the number of events in the reconstruction level $y_W$ bin $i$.
The migration matrices provide information about the relation between events selected at 
reconstruction level and the original events at generator level and are used to correct the 
data for detector resolution effects. The procedure is validated by using events 
generated with {\sc MC@NLO}, where we find good agreement between the unfolded and the 
generated $W$  boson charge asymmetry.

The primary systematic uncertainties on asymmetry come from the unfolding procedure 
including the uncertainties from the event migration correction, 
the acceptance and efficiency correction, and the PDF inputs (fractional uncertainty, [1.1--5.0]$\times 10^{-3}$). 
To estimate the uncertainty from the PDF inputs,
we determine the $Q\left(y_W, p_T^W\right)$ correction with 45 {\sc CTEQ6.6} PDF sets,
perform the measurement with different $Q\left(y_W, p_T^W\right)$~\cite{supplemental},
and extract the uncertainty for each $y_W$ bin using the prescription described in Ref.~\cite{cteq}.
Other systematic uncertainties arise from the modeling of the
$p_T^W$ distribution and the final state radiation modeling ([0.1--2.4]$\times10^{-4}$), electron 
identification corrections ([0.1--0.7]$\times 10^{-3}$), electron energy modeling ([0.1--0.5]$\times 10^{-3}$), 
hadronic recoil modeling ([0.1--0.8]$\times 10^{-3}$), background modeling ([0.1--1.0]$\times 10^{-3}$), 
MC modeling imperfections ([0.2--2.6]$\times 10^{-3}$), electron charge misidentification ([0.1--2.0]$\times 10^{-3}$),
and the relative efficiency for positrons and
electrons ($K^{\pm}_{\text{eff}}$) ([0.1--0.6]$\times 10^{-3}$). 
 
\begin{center}
\begin{figure}
\epsfig{file=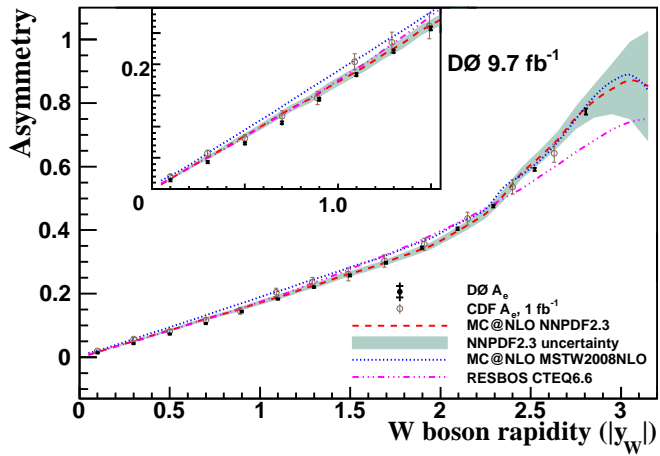, scale=0.48}
\caption{\small (color online). Measured $W$ boson charge asymmetry, after {\it CP}
folding, compared to predictions and the CDF 1 fb$^{-1}$ result.
The points show the measured asymmetry, with the horizontal bars
delineating the
statistical uncertainty component and the vertical lines showing the total uncertainty.
The central value and uncertainty from {\sc MC@NLO} using {\sc NNPDF2.3} PDF sets and the prediction from
{\sc resbos} using the {\sc CTEQ6.6} central PDF set, {\sc MC@NLO} using the {\sc MSTW2008NLO} 
central PDF set are also shown.
The inset focuses on the $y_W$ region from 0 to 1.5.} 
\label{fig:compare_asym_old}
\end{figure}
\end{center}

Figure~\ref{fig:compare_asym_old} shows the measured values of the $W$ boson asymmetry
together with the result from CDF~\cite{cdf_w}. The data are compared to the {\sc MC@NLO} prediction 
with the {\sc NNPDF2.3}~\cite{nnpdf} PDF set, 
next-to-leading order {\sc resbos} prediction with {\sc photos}~\cite{photos} using
the {\sc CTEQ6.6} central PDF set, and {\sc MC@NLO} using {\sc MSTW2008NLO}~\cite{mstw} central PDF set.
In the predictions, we require both the electron
and neutrino transverse momentum to be above $25$~GeV and merge the radiated
photons into the electron
if they fall within a cone of radius $\Delta R = \sqrt{(\Delta \phi)^2 +
(\Delta \eta)^2} < 0.3$.
There is agreement between the data and predictions, although the predictions are systematically higher
than the data by $\sim$ 1 standard deviation in all measurements for $|y_W|$ between 0.1 and 1.
Values of the asymmetry in bins of $y_W$, average bin positions, and predictions are shown in
Table~\ref{Tab:Data_w_fold_asym1_old}.
The experimental uncertainties are substantially smaller than the uncertainties from the
{\sc NNPDF2.3} PDF sets in all $y_W$ bins, demonstrating the importance
of this analysis to improve PDFs.
Table~\ref{Tab:Correlation_All_Fold_Type_W} lists the correlations
between central values in different $y_W$ bins that are introduced by the ambiguity in $p_z^{\nu}$.
The correlation coefficients of systematic uncertainties between different $y_W$ are negligible. 

\begin{table}
\begin{center}
\caption{{\it CP}-folded $W$ charge asymmetry for data and predictions
from {\sc MC@NLO} using {\sc NNPDF2.3} PDFs tabulated in percent (\%)
for each $|y_W|$ bin. The $\langle|y_W|\rangle$ is calculated as the cross section weighted average of $y_W$ in each 
bin from {\sc resbos} with {\sc photos}. For data, the first uncertainty is statistical and 
the second is systematic. 
The uncertainties on the prediction come from both the PDF uncertainties
and $\alpha_s$ uncertainties. }
\begin{tabular}{cccrr}
\hline
\hline
 Bin index &  $|y_W|$ &  $\langle|y_W|\rangle$ &   \multicolumn{1}{c}{Data} & Prediction      \\ \hline 
\ \ 1&  0.0--0.2 &$0.10$ & $1.40 \pm 0.17 \pm 0.12$  &  $1.61 \pm 0.19$  \\ 
\ \ 2&  0.2--0.4 &$0.30$ & $4.32 \pm 0.18 \pm 0.19$  &  $5.06 \pm 0.33$ \\ 
\ \ 3&  0.4--0.6 &$0.50$ & $7.33 \pm 0.19 \pm 0.27$  &  $8.50 \pm 0.41$  \\ 
\ \ 4&  0.6--0.8 &$0.70$ & $10.59 \pm 0.20 \pm 0.32$  &  $12.05 \pm 0.53$  \\ 
\ \ 5&  0.8--1.0 &$0.90$ & $14.36 \pm 0.21 \pm 0.34$ &  $15.36 \pm 0.66$ \\ 
\ \ 6&  1.0--1.2 &$1.10$ & $18.32 \pm 0.22 \pm 0.37$ &  $18.86 \pm 0.74$ \\ 
\ \ 7&  1.2--1.4 &$1.30$ & $22.06 \pm 0.24 \pm 0.39$ &  $22.52 \pm 0.80$   \\ 
\ \ 8&  1.4--1.6 &$1.50$ & $25.74 \pm 0.27 \pm 0.36$ &  $26.30 \pm 0.85$   \\ 
\ \ 9& 1.6--1.8 &$1.70$ & $29.75 \pm 0.31 \pm 0.34$ &  $29.89 \pm 0.92$   \\ 
10 &  1.8--2.0 &$1.90$ & $34.46 \pm 0.35 \pm 0.38$   &  $34.04 \pm 1.08$  \\ 
11 &  2.0--2.2 &$2.10$ & $40.42 \pm 0.40 \pm 0.43$   &  $39.77 \pm 1.31$ \\ 
12 &  2.2--2.4 &$2.29$ & $47.55 \pm 0.44 \pm 0.43$   &  $47.73 \pm 1.62$  \\ 
13 &  2.4--2.7 &$2.52$ & $59.10 \pm 0.46 \pm 0.44$   &  $61.81 \pm 1.74$  \\ 
14 &  2.7--3.2 &$2.81$ & $77.33 \pm 0.93 \pm 0.56$   &  $78.05 \pm 4.36$  \\ 
\hline
\hline
\end{tabular}
\label{Tab:Data_w_fold_asym1_old}
\end{center}
\end{table}

\begin{table*}
\begin{center}
\caption{Correlation coefficients between central values of asymmetry in different $|y_W|$ bins.}
\begin{tabular}{c|cccccccccccccc}
\hline
\hline
$|y_W|$ bin & 1    & 2    & 3    & 4    & 5    & 6    & 7    & 8    & 9    & 10   & 11   & 12   & 13  & 14 \\ \hline
 
1 & 1.00 & 0.84 & 0.57 & 0.38 & 0.29 & 0.25 & 0.21 & 0.16 & 0.10 & 0.06 & 0.04 & 0.03 & 0.02 & 0.01 \\
2 &  & 1.00 & 0.85 & 0.58 & 0.39 & 0.29 & 0.24 & 0.16 & 0.11 & 0.07 & 0.04 & 0.04 & 0.03 & 0.02 \\
3 &  &  & 1.00 & 0.85 & 0.58 & 0.38 & 0.26 & 0.16 & 0.10 & 0.06 & 0.05 & 0.06 & 0.05 & 0.03 \\
4 &  &  &  & 1.00 & 0.83 & 0.52 & 0.29 & 0.16 & 0.09 & 0.07 & 0.08 & 0.10 & 0.09 & 0.06 \\
5 &  &  &  &  & 1.00 & 0.78 & 0.42 & 0.19 & 0.11 & 0.10 & 0.13 & 0.15 & 0.14 & 0.10 \\
6 &  &  &  &  &  & 1.00 & 0.74 & 0.37 & 0.22 & 0.19 & 0.22 & 0.22 & 0.20 & 0.15 \\
7 &  &  &  &  &  &  & 1.00 & 0.76 & 0.50 & 0.38 & 0.34 & 0.31 & 0.29 & 0.21 \\
8 &  &  &  &  &  &  &  & 1.00 & 0.84 & 0.62 & 0.47 & 0.38 & 0.34 & 0.27 \\
9 &  &  &  &  &  &  &  &  & 1.00 & 0.87 & 0.65 & 0.48 & 0.40 & 0.31 \\
10 &  &  &  &  &  &  &  &  &  & 1.00 & 0.89 & 0.67 & 0.51 & 0.36 \\
11 &  &  &  &  &  &  &  &  &  &  & 1.00 & 0.89 & 0.66 & 0.41 \\
12 &  &  &  &  &  &  &  &  &  &  &  & 1.00 & 0.86 & 0.45 \\
13 &  &  &  &  &  &  &  &  &  &  &  &  & 1.00 & 0.50 \\
14 &  &  &  &  &  &  &  &  &  &  &  &  &  & 1.00 \\
\hline
\hline
\end{tabular}
\label{Tab:Correlation_All_Fold_Type_W}
\end{center}
\end{table*}

\indent In summary, we have measured the $W$ boson charge asymmetry in $p\bar{p}\rightarrow W \rightarrow e\nu$
events by using data corresponding to 9.7~fb$^{-1}$ of integrated luminosity collected by the D0 experiment at $\sqrt{s}=1.96$~TeV. 
By using the neutrino weighting method, the most precise direct measurement of the $W$ boson charge asymmetry to date is obtained. 
With coverage extended to $|\eta^e|=3.2$,
this measurement can be used to improve the precision and accuracy of next-generation PDF sets;
in particular, it provides more accurate information for PDFs at high $x$, compared with
measurements of the lepton charge asymmetry, which is crucial for many beyond SM searches.

\section{Acknowledgements}
%
We thank the staffs at Fermilab and collaborating institutions
and acknowledge support from the
DOE and NSF (USA);
CEA and CNRS/IN2P3 (France);
MON, NRC KI, and RFBR (Russia);
CNPq, FAPERJ, FAPESP, and FUNDUNESP (Brazil);
DAE and DST (India);
Colciencias (Colombia);
CONACyT (Mexico);
NRF (Korea);
FOM (The Netherlands);
STFC and the Royal Society (United Kingdom);
MSMT and GACR (Czech Republic);
BMBF and DFG (Germany);
SFI (Ireland);
The Swedish Research Council (Sweden);
and
CAS and CNSF (China).
%


\begin{thebibliography}{99}
\vskip 0.25cm

\bibitem{d0_w}
 V. M. Abazov {\it et al.} (D0 Collaboration), Phys. Rev. Lett {\bf 112}, 151803 (2014).

\bibitem{w_prd}
 V. M. Abazov {\it et al.} (D0 Collaboration), arXiv: 1412.2862 (2014), submitted to Phys. Rev. D.

\bibitem{cdf_w}
 T. Aaltonen {\it et al.} (CDF Collaboration), Phys. Rev. Lett. {\bf 102}, 181801 (2009).

 \bibitem{mcatnlo}
S. Frixione and B. R. Webber, J. High Energy Phys. {\bf 06}, 029 (2002).


\bibitem{nnpdf}
R. D. Ball {\it et al.}, Nucl. Phys. {\bf B867}, 244 (2013).

\bibitem{resbos}
 C. Balazs and C. P. Yuan, Phys. Rev. D {\bf 56}, 5558 (1997).

 \bibitem{cteq}
 J. Pumplin, D. R. Stump, J. Huston, H.-L. Lai, P. Nadolsky, and W.-K. Tung, J. High Energy Phys. {\bf 07} 012 (2002);
D. Stump, J. Huston, J. Pumplin, W.-K Tung, H.-L. Lai, S. Kuhlmann, and J. F. Ownes, J. High Energy Phys. {\bf 10}, 046 (2003).

 \bibitem{mstw}
 A.D. Martin, W. J. Stirling, R. S. Thorne, and G. Watt, Eur. Phys. J. C {\bf 63}, 189 (2009).


 \bibitem{photos}
P. Golonka and Z. W$\text{\c{a}}$s, Eur. Phys. J. C {\bf 45}, 97 (2006).



\end{thebibliography}

\begin{thebibliography}{99}
\vskip 0.25cm
 \bibitem{old_pdf1}
 E.L. Berger, F. Halzen, C.S. Kim and S. Willenbrock, Phys. Rev. D {\bf 40}, 83 (1989).
 
  \bibitem{old_pdf2}
A.D. Martin, R.G. Roberts, and W.J. Stirling, Mod. Phys. Lett. A {\bf 4}, 1135 (1989).

\bibitem{old_pdf3}
H.-L. Lai, J. Boots, J. Huston, J. G. Morfin, J. F. Owens, J. W. Qiu, W.-K. Tung, and H. Weerts, Phys. Rev. D {\bf 51}, 4763 (1995).

\bibitem{CDF_results1}
F. Abe {\it et al.} (CDF Collaboration), Phys. Rev. Lett. {\bf 74}, 850 (1995).

\bibitem{CDF_results2}
F. Abe {\it et al.} (CDF Collaboration), Phys. Rev. Lett. {\bf 81}, 5754 (1998).

\bibitem{CDF_results3}
D. Acosta {\it et al.} (CDF Collaboration), Phys. Rev. D {\bf 71}, 051104 (2005).

\bibitem{d0_results_muon1}
V. M. Abazov {\it et al.} (D0 Collaboration), Phys. Rev. D {\bf 77}, 011106 (2008).

\bibitem{d0_results_em_old}
V. M. Abazov {\it et al.} (D0 Collaboration), Phys. Rev. Lett. {\bf 101}, 211801 (2008).

 \bibitem{d0_muon}
 V. M. Abazov {\it et al.} (D0 Collaboration), Phys. Rev. D {\bf 88}, 091102(R) (2013).

\bibitem{ATLAS_w}
G. Aad {\it et al.} (ATLAS Collaboration), Phys. Lett. B {\bf 701}, 31 (2011).

\bibitem{CMS_w}
S. Chatrchyan {\it et al.} (CMS Collaboration), Phys. Rev. Lett. {\bf 109}, 111806 (2012).

 \bibitem{cdf_w}
 T. Aaltonen {\it et al.} (CDF Collaboration), Phys. Rev. Lett. {\bf 102}, 181801 (2009).

 \bibitem{cdf_method}
 A. Bodek, Y. Chung, B.-Y. Han, K. McFarland, and E. Halkiadakis, Phys. Rev. D {\bf 77}, 111301(R) (2008).

\bibitem{d0lumi}
 T.~Andeen {\it et al.}, Report No. FERMILAB-TM-2365, 2007.

\bibitem{d0det0}
 S. Abachi {\it et al.} (D0 Collaboration), Nucl. Instrum. Methods Phys. Res., Sect. A {\bf 338}, 185 (1994).

\bibitem{d0det}
 V. M. Abazov {\it et al.} (D0 Collaboration), Nucl. Instrum. Methods Phys. Res., Sect. A {\bf 565}, 463 (2006).

\bibitem{d0_coordinate}
D0 uses a cylindrical coordinate system with the $z$ axis
along the beam axis in the proton direction. Angles $\theta$ and $\phi$ are the polar and
azimuthal angles, respectively. Pseudorapidity is defined as
$\eta=-\ln [\tan(\theta/2)]$ where $\theta$ is measured with respect
to the interaction vertex. In the massless limit, $\eta$ is equivalent to
the rapidity $y=(1/2) \ln[(E+p_z)/(E-p_z)]$,
and $\eta_{\text{det}}$ is the pseudorapidity measured with respect to the
center of the detector. 

\bibitem{shower_shape}
S. Abachi {\it et al.} (D0 Collaboration), Nucl. Instrum. Meth. Phys. Res., Sect. A {\bf 324}, 53 (1993).

 \bibitem{tag-probe}
 V. Abazov {\it et al.} (D0 Collaboration), Phys. Rev. D {\bf 76}, 012003 (2007).

\bibitem{pythia}
 T. Sj$\ddot{\text o}$strand, P. Ed$\acute{\text{e}}$n, C. Feriberg, L. L$\ddot{\text o}$nnblad, G. Miu, S. Mrenna, and E. Norrbin, Comput. Phys. Commun. {\bf 135}, 238 (2001). {\sc pythia} version v6.323 is used.

\bibitem{cteq}
  J. Pumplin, D. R. Stump, J. Huston, H.-L. Lai, P. Nadolsky, and W.-K. Tung, J. High Energy Phys. 07 (2002) 012; 
  D. Stump, J. Huston, J. Pumplin, W.-K Tung, H.-L. Lai, S. Kuhlmann, and J. F. Ownes, J. High Energy Phys. 10 (2003) 046.

\bibitem{geant}
  R. Brun and F. Carminati, CERN Program Library Long Writeup
  W5013, 1993 (unpublished).

\bibitem{resbos}
  C. Balazs and C. P. Yuan, Phys. Rev. D {\bf 56}, 5558 (1997).

\bibitem{nlo_xsection}
 R. Hamberg, W.L. van Neerven, and T. Matsuura, Nucl. Phys. B {\bf 359}, 343 (1991).

\bibitem{wmass}
J. Beringer {\it et al.} (Particle Data Group), Phys. Rev. D {\bf 86}, 010001 (2012).

\bibitem{mcatnlo}
S. Frixione and B. R. Webber, J. High Energy Phys. 06 (2002) 029.

\bibitem{vrap}
C. Anastasiou, L. Dixon, K. Melnikov, and F. Petriello, Phys. Rev. D
{\bf 69}, 094008 (2004).

\bibitem{matrix}
  G. L. Marchuk, {\it Methods of Numerical Mathematics} (Springer, Berlin, 1975).
  
\bibitem{supplemental}
  See Supplemental Material at http://link.aps.org/supplemental/10.1103/PhysRevLett.112.151803
  for the measured $W$ boson charge asymmetry central values using 45
  {\sc CTEQ6.6} different PDF sets as input.

\bibitem{nnpdf}
  R. D. Ball {\it et al.}, Nucl. Phys. B {\bf 867}, 244 (2013).

\bibitem{photos}
P. Golonka and Z. W\c{a}s, Eur. Phys. J. C {\bf 45}, 97 (2006).

 \bibitem{mstw}
 A.D. Martin, W. J. Stirling, R. S. Thorne, and G. Watt, Eur. Phys. J. C {\bf 63}, 189 (2009).
\end{thebibliography}
\end{document}